\documentclass{iaus}
\usepackage{graphicx}

\title[The Circumstellar Discs of Be Stars] 
{The Circumstellar Discs of Be Stars}

\author[Carciofi, A. C.]   
{Alex C. Carciofi$^1$}

\affiliation{$^1$
Instituto de Astronomia, Geof\'{i}sica e Ci\^encias Atmosf\'ericas, \\ Universidade de S\~ao Paulo, 
Rua do Mat\~ao 1226, Cidade Universit\'aria, \\05508-900, S\~ao Paulo, SP, BRAZIL 
email: {\tt carciofi@usp.br}
}

\pubyear{2010}
\volume{272}  
\pagerange{001-012}
\jname{Active OB Stars: structure, evolution, mass loss}
\editors{C. Neiner, G. Wade, G. Meynet \& G. Peters, eds.}
\begin{document}

\maketitle

\begin{abstract}
Circumstellar discs of Be stars are thought to be formed from material ejected from a fast-spinning central star. This material possesses large amounts of angular momentum and settles in a quasi-Keplerian orbit around the star. This simple description outlines the basic issues that a successful disc theory must address: 
1) What is the mechanism responsible for the mass ejection? 
2) What is the final configuration of the material? 
3) How the disc grows? 
With the very high angular resolution that can be achieved with modern interferometers operating in the optical and infrared we can now resolve the photosphere and immediate vicinity of nearby Be stars. Those observations are able to provide very stringent tests for our ideas about the physical processes operating in those objects. 
This paper discusses the basic hydrodynamics of viscous decretion discs around Be stars. The model predictions are quantitatively compared to observations, demonstrating that the viscous decretion scenario is currently the most viable theory to explain the discs around Be stars.
\end{abstract}

\firstsection 
\section{Introduction}

Recent years witnessed an important progress in our understanding of the circumstellar discs of Be stars, largely due to interferometric observations capable of angularly resolve those objects at the milliarcsecond (mas) level (see \cite[Stee 2010]{ste10}, for a review).

In the late nineties, discs around Be stars were considered to be equatorially enhanced outflowing winds, and several models and mechanisms to drive the outflow were proposed (see \cite[Bjorkman 2000]{bjo00}, for a review).
The much stronger observational constraints available today allow us to rule out several theoretical scenarios that were proposed in the past (e.g. the wind compressed disc models of \cite[Bjorkman \& Cassinelli 1993]{bjo93}). As an example, spectrointerfometry and spectroastrometry have directly probed the disc kinematics (\cite[Meilland et al. 2007b]{mei07a}, \cite[\v{S}tefl et al. 2010]{stefl10}, \cite[Oudmaijer et al. 2010]{oud10}), revealing that Be discs rotate very close to Keplerian.
As a result, the viscous decretion scenario, proposed originally by \cite{lee91} and further developed by \cite{por99}, \cite{oka01}, \cite{bjo05}, among others, has emerged as the most viable scenario to explain the observed properties of Be discs.


\section{Discs Diagnostics \label{observations}}

Before reviewing the basic aspects of the theory of circumstellar discs, it is useful to put in perspective what observations tells us about the structure and kinematics of those discs. 

\subsection{The formation loci of different observables}

\begin{figure}[t]
\begin{center}
 \includegraphics[width=2.6in]{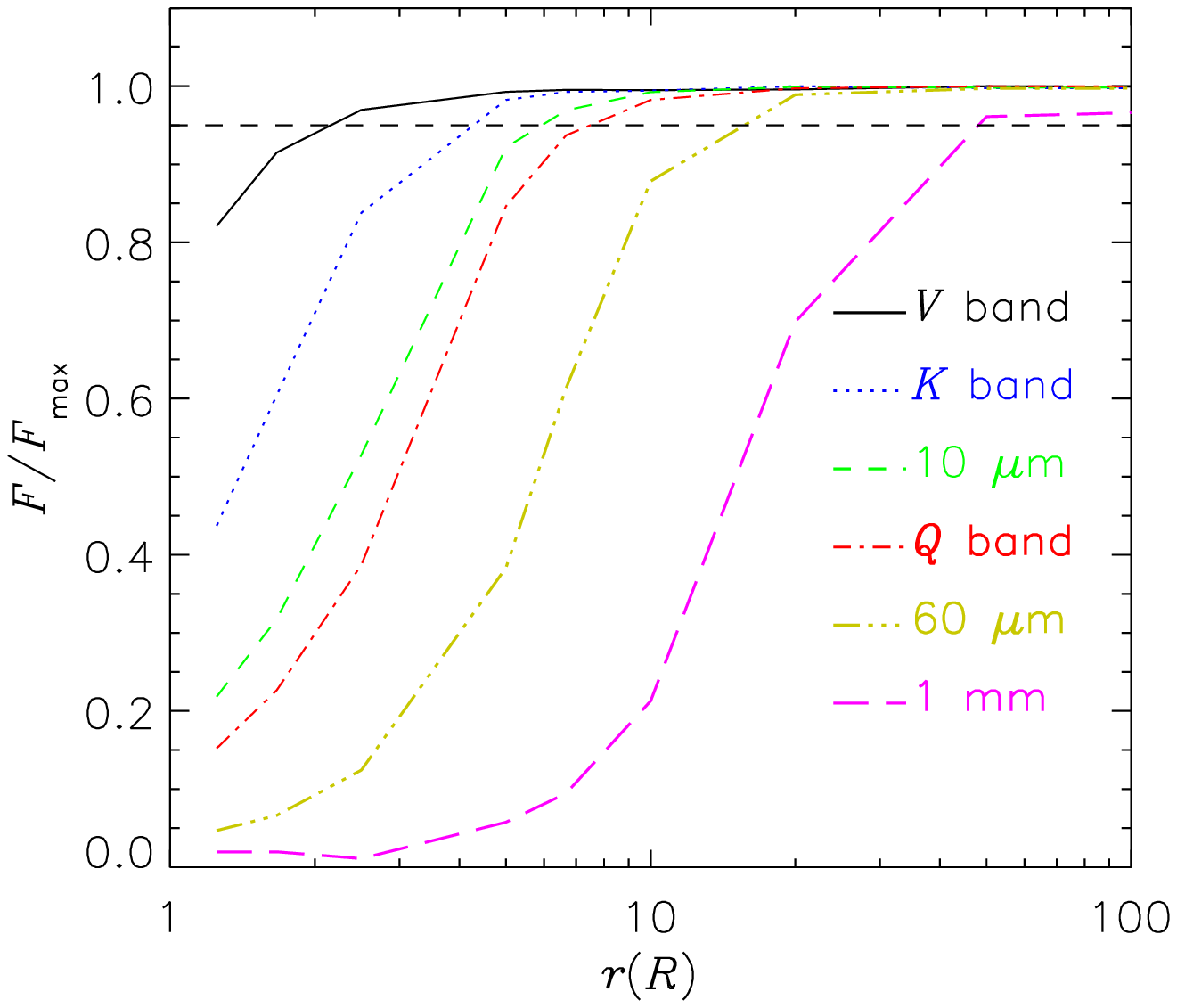} 
 \includegraphics[width=2.6in]{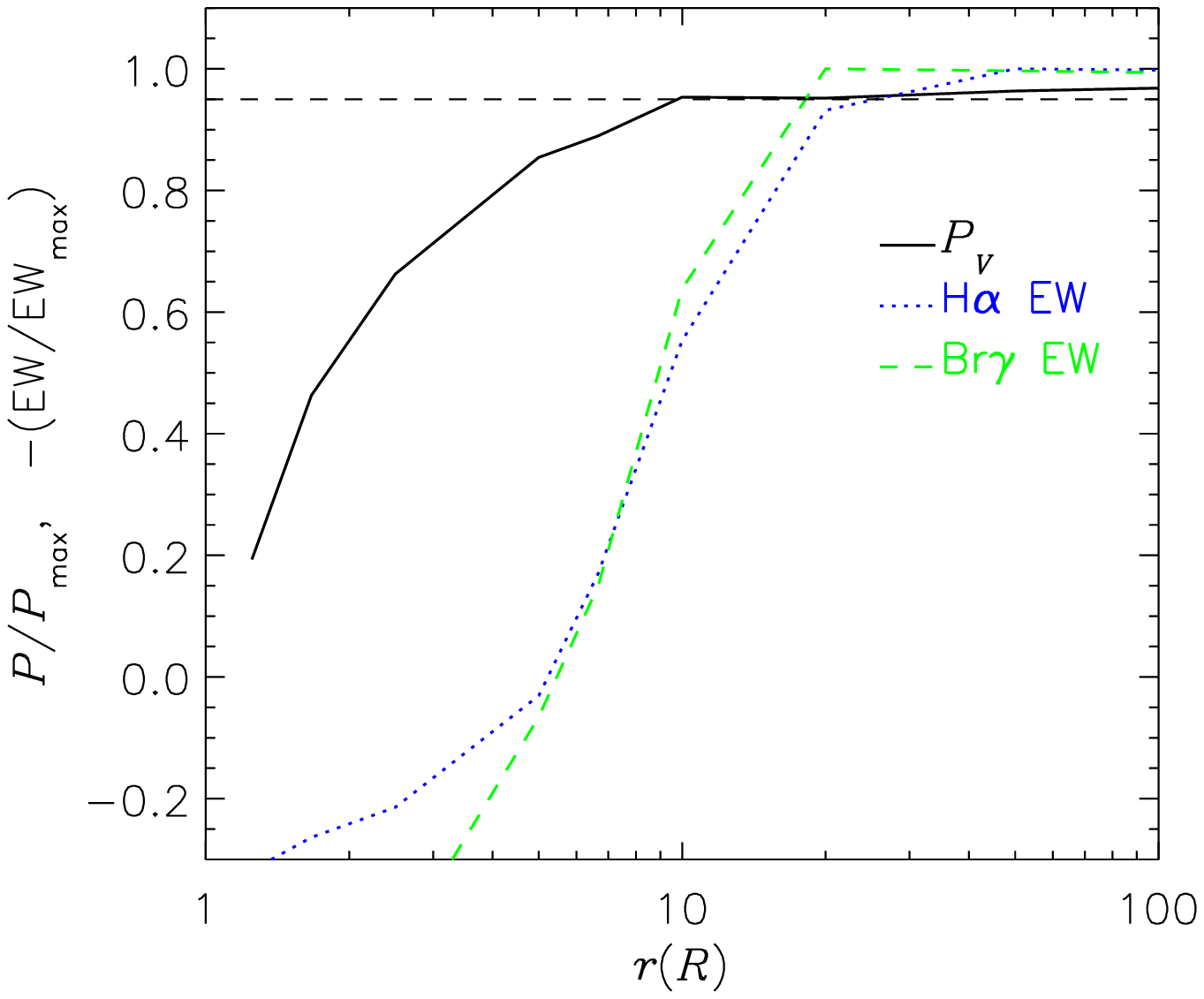} 
 \caption{The formation loci of different observables. The calculations assume a rapidly rotating B1Ve star ($T_{\rm eff}^{\rm pole} = 25\,000\;\rm K$, $\Omega/\Omega_{\rm crit} = 0.92$) surrounded by viscous decretion discs with different sizes. The results correspond to a viewing angle of $30^\circ$.
{{\it Left: continuum emission}}.  Plotted is the ratio between the observed flux to the maximum flux, $F_{\rm max}$, as a function of the disc outer radius. $F_{\rm max}$ corresponds to the flux of a model with a disc outer radius of $1000\;\rm R$.
{{\it Right: polarization and line equivalent width}}.  
Calculations were carried out with the {\sc hdust} code (\cite[Carciofi \& Bjorkman 2006]{car06a}). 
  }
   \label{formation_loci}
\end{center}
\end{figure}

One important issue to consider when analyzing observations is that  different observables probe different regions of the disc; it is therefore useful to be able to make a correspondence between a given observable (say, the continuum flux level at a given wavelength) and the part of the disc whence it comes.

To make such a correspondence, we consider a typical Be star of spectral type B1Ve and calculate the emergent spectrum arising from viscous decretion discs of different outer radii (see Sect.~\ref{viscous_discs} for a detailed description of the structure of viscous discs). In Fig.~\ref{formation_loci} we present results for continuum emission, continuum polarization and the equivalent width of emission lines.

Let us discuss first the continuum and line emission. The reason the spectra of Be stars show continuum excess or emission lines is that the dense disc material acts as a pseudo-photosphere that is much larger than the stellar photosphere. For the pole-on case (inclination $i=0$), the effective radius of the pseudo-photosphere is the location where the vertical optical depth $\tau_\lambda(R_{\rm eff}) = 1$. For continuum Hydrogen absorption, the opacity depends on the wavelength as $\kappa_\lambda \propto \lambda^{2}$ (if one neglects the wavelength dependence of the absorption gaunt factors) and thus the size of the pseudo-photosphere increases with wavelength (see \cite[Carciofi \& Bjorkman 2006]{car06a}, appendix A, for a derivation of $R_{\rm eff}(\lambda)$). 

In Fig.~\ref{formation_loci}, left panel, the intersection of the different lines (each corresponding to a given wavelength) with the horizontal dashed line marks the position in the disc whence about 95\% of the continuum excess comes. For instance, from the Figure we see that the $V$ band excess is formed very close to the star, within about $2\;R$, whereas the excess at 1 mm originates from a much larger volume of the disc. As we shall see below, the fact that the continuum excess flux at visible wavelengths forms so close to the star makes such observations indispensable to study systems with complex temporal evolution because the V band continuum emission has the fastest response to disc changes as a result of photospheric activity.

In the right panel of Fig.~\ref{formation_loci} we show how the line emission and continuum polarization grows with radius.  For the optically thick H$\alpha$ and Br$\gamma$ lines the disc emission only fills in the photospheric absoprtion profile when the disc size is about $5\;R$. Both lines have pseudo-photospheres that extend up to about $20\;R$.
For polarization, the results are at odds with the common belief that polarization is formed close to the star; it is seen that 95\% of the maximum polarization is only reached when the disc size is about $10\;R$.


\subsection{Disc Thickness}

In a nice example of the diagnostics potential of interferometric and polarimetric observations combined, \cite{woo97} and \cite{qui97} showed that the circumstellar disc of $\zeta$ Tau is geometrically thin (opening angle of $2.5^\circ$). This result was confirmed by \cite{car09} from a more detailed analysis. Other studies based on the fraction of Be-shell stars vs. Be stars typically find larger values for the opening angle (e.g. $13^\circ$ from \cite[Hanuschik 1996]{han96}) but this discrepancy is accommodated by the fact that the geometrical thickness of Be discs increases with radius (flared disc) and different observables probe different disc regions (Fig.~\ref{formation_loci}). In any case, there is no doubt that the Be discs are flat, geometrically thin structures.


\subsection{Disc density distribution \label{density}}

Models derived both empirically (\cite[e.g., Waters 1986]{wat86}) and theoretically (e.g., \cite[Okazaki 2001]{oka01} and \cite[Bjorkman \& Carciofi 2005]{bjo05}) usually predict (or assume) a power-law fall off of the disc density [$\rho(r) \propto r^{-n}]$.  Values for $n$ vary widely in the literature, typically in the range $2<n<4$.
One question that arises is whether this quoted range for the density slope is real or not. 

Below we see that the viscous decretion disc model, in its simplest form of an isothermal and isolated disc, predicts a slope of $n=3.5$ for the disc density. On one hand, the inclusion of other physical effects can change the value of the above slope. For instance, non-isothermal viscous diffusion results in much more complex density distribution that cannot be well represented by a simple power-law, and tidal effects by a close binary may make the density slope shallower (\cite[Okazaki et al. 2002]{oka02}). On the other hand, it is also true that many of quoted values for $n$ in the literature are heavily influenced by the assumptions and methods used in the analysis and should be viewed with caution. 

\subsection{Disc dynamics}

Since the earliest detections of Be stars, it became clear that a rotating circumstellar disc-like material was the most natural explanation for the observed double-peaked profile of emission lines, but the problem of how the disc rotates has been an open issue until recently. 

The disc rotation law is profoundly linked with the disc formation mechanism; therefore, determining observationally the disc dynamics is of great interest. Typically, one can envisage three limiting cases for the radial dependence of the azimuthal component of the velocity, $v_\phi$, depending on the forces acting on the disc material
\begin{equation}
v_{\phi} = \left\{
 \begin{array}{ll}
V_{\rm rot}(r/R)^{-1} & \mbox{radiatively driven outflow}\,, \\
V_\star(r/R) & \mbox{magnetically dominated disc} \,, \\
V_{\rm crit}(R/r)^{1/2} & \mbox{disc driven by viscosity}\,.\\
\end{array}
\right.
\end{equation}
In the first case (radiatively driven outflow) the dominant force on the material is the radially directed radiation pressure that does not exert torques and thus conserves angular momentum ($V_{\rm rot}$ is thus the rotation velocity of the material when it left the stellar surface).
The second case corresponds to a rigidly rotating magnestosphere a la $\sigma$ Ori E (\cite[Townsend et al. 2005]{tow05}), in which the plasma is forced to rotate at the same speed as the magnetic field lines ($V_\star$).
The last case corresponds to Keplerian orbital rotation, written in terms of the critical velocity, $V_{\rm crit} \equiv \left(GM/R\right)^{1/2}$, which is the Keplerian orbital speed at the stellar surface. This case  requires a fine-tunning mechanism such that \emph{the centrifugal force, $v_\phi^2/r$,  exactly balances gravity at all radii}. As we shall see below, viscosity does provide the fine-tunning mechanism capable of producing a Keplerian disc.

\cite{por03} reviewed the then existing observational constraints on the disc kinematics and concluded that ''all of the kinematic evidence seems to point to a disc velocity field dominated by rotation, with little or no radial flow, at least in the regions where the kinematic signatures of emission and absorption are significant''. Today, spectrointerferometry and spectroastrometry provides clear-cut evidence that, in most systems observed and analysed so far ($\kappa$ CMa being the only possible exception, \cite[Meilland et al. 2007b]{mei07b}), the discs rotate in a Keplerian fashion (\cite[Meilland et al. 2007b]{mei07a}, \cite[\v{S}tefl et al. 2010]{stefl10}, \cite[Oudmaijer et al. 2010]{oud10}). 
This is an important result, seeing that it indicates that viscosity is the driving mechanism of the outflow.

\subsection{Cyclic $V/R$ variations}

About 2/3 of the Be stars present the so-called $V/R$ variations, a phenomenon characterised by the quasi-cyclic variation in the ratio between the violet and red emission peaks of the H~I emission lines. These variations are generally explained by global oscillations in the circumstellar disc forming a one-armed spiral density pattern that precesses around the star with a period of a few years (\cite[Kato 1983]{kat83}, \cite[Okazaki 1991]{oka91}, \cite[Okazaki 1997]{oka97}). 

Recently, \cite{car09} provided a quantitative verification of the global disc oscillation theory from a detailed modelling of high-angular resolution {\sc amber} data of the Be star $\zeta$ Tau. From a theoretical perspective, the existence of density waves in rotating discs, as suggested by \cite{kat83}, imposes most stringent constraints on the rotation velocity, since mode confinement requires that the rotation law be Keplerian within about 1\% and the radial flow be hightly subsonic ($\lesssim 0.01\;c_s$, \cite[Okazaki 2007]{oka07}).

\subsection{Long-term variations}

\begin{figure}[t]
\begin{center}
 \includegraphics[width=2.0in, angle=90, viewport=00 50 362 750]{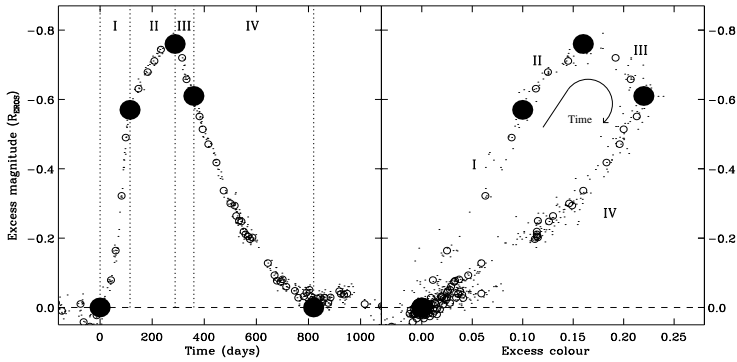} 
 \caption{ Light curve ({\it left panel}) and CMD ({\it right panel}) of the variation of the star OGLE 005209.92-731820.4 (\cite[de Wit 2006]{dew06}, reproduced with permission).
  }
   \label{dewit}
\end{center}
\end{figure}

One of the most intriguing types of variability observed in the Be stars is the aperiodic transition between a normal B phase (discless phase) and a Be phase whereby the disc is lost and rebuilt in timescales of months to years (e.g., \cite[Clark et al. 2003]{cla03}, \cite[\v{S}tefl et al. 2010]{stefl10}).
The varying amount of circumstellar gas manifests itself as changes in line profiles (\cite[Clark et al. 2003]{cla03}), continuum brightness and colors (\cite[Harmanec 1983]{har83}) and polarization (\cite[Draper et al. 2010]{dra10}).  
A fine example of the secular process of disc formation and dissipation, and its effects on the continuum brightness and colors, is shown in Fig.~\ref{dewit}: the outburst phase responsible for the disc build-up lasted about 300 hundred days (phases I and II, during which the star got brighter and redder) and was followed by a quiescent phase of about 500 days (phases III and IV, during which the star slowly went back to its original appearance as the previously built disc dissipated).
As discussed below, the timescales involved in the disc formation and dissipation are generally consistent with the timescales of viscous diffusion.

\subsection{Short-term variations}

Small-scale, short-term variations are quite common in Be stars, and possess a complex phenomenology (\cite[see Rivinius 2007, for a review]{riv07}). On one hand, some observed variations (e.g. line profile variations due to non-radial pulsations) are associated with the photosphere proper; others, on the other hand, are thought to originate from the very base of the disc and are, therefore, the manifestation of the physical process(es) that is (are) feeding the disc (e.g. short-term $V/R$ variations of emission lines).
To date, $\mu$ Cen remains the only system in which the ejection mechanism have been unambiguously identified (in this case non-radial pulsations, \cite[Rivinius et al. 1998]{riv98}). 
In the viscous diffusion theory outlined below, it is assumed that matter is ejected by the star and deposited at the inner boundary of the disc with Keplerian or super-Keplerian speeds; clearly, the current status of the theory is still unsatisfactory inasmuch as the fundamental link between the disc and the photosphere proper (i.e., the feeding mechanism) is still unknown.

Studies of short-term variations associated with the disc (e.g., \cite[Rivinius et al. 1998]{riv98}) are generally consistent with the following scenario: mass injection by some mechanism that causes a transient asymmetry in the inner disc that manifests itself by, e.g., short-term $V/R$ variations, followed by circularization and dissipation in a viscous scale (see below). Here, again, viscosity seems to be playing a major role in shaping the evolution of ejecta.

\section{Viscous Decretion Disc Models  \label{viscous_discs}}

The observational properties outlined above give strong hints as to the ingredients a successful theory for the structure of the Be discs must have. The theory must explain the keplerian rotation, slow outflow speeds, small geometrical thickness and must also account for the timescales of disc build-up and dissipation.

The only theory to date that satisfies those requirements is the viscous decretion disc model (\cite[Lee et al. 1991]{lee91}). This model is essentially the same as that employed for protostellar 
discs (\cite[Pringle 1981]{pri81}), the primary difference being that Be discs are outßowing, while pre- 
main-sequence discs are inflowing. In this model, it is supposed that some yet unknown mechanism injects material at the Keplerian orbital speed
into the base of the disc. Eddy/turbulent viscosity then transports angular 
momentum outward from the inner boundary of the disc (note that this requires 
a continual injection of angular momentum into the base of the disc). If the 
radial density gradient is steep enough, angular momentum is added 
to the individual fluid elements and they slowly move outward. To critically test 
such decretion disc models of Be stars against observations, we must determine 
the structure of the disc from hydrodynamical considerations.

\subsection{Viscous discs fed by constant decretion rates}

There are many solutions available for the case of a viscous disc fed by a constant decretion rate, all agreing in their essentials. Below we outline the mains steps of one possible derivation (\cite[Bjorkman 1997]{bjo97}) to the goal of discussing the properties of the solution and confronting them to the observations. More detailed analyses as well as different approaches to the problem can be found in \cite{lee91}, \cite{oka01}, \cite{bjo05}, \cite{oka07}, \cite{jon08} and \cite{car08}.

\subsection{Hydrostatic Structure} 

Our goal is to write and solve the Navier Stokes fluid equations in cylindrical coordinates ($\varpi,\phi,z$) which in the steady-state case have the following general form
\begin{eqnarray}
   \frac{1}{\varpi}\frac{\partial}{\partial\varpi}(\varpi\rho v_\varpi)
 + \frac{1}{\varpi}\frac{\partial}{\partial\phi}  (\rho v_\phi)
 +                 \frac{\partial}{\partial z}    (\rho v_z)
  &=& 0 \enspace , \label{eq:continuity} \\
   v_\varpi \frac{\partial v_\varpi}{\partial \varpi}  
 + \frac{v_\phi}{\varpi} \frac{\partial v_\varpi}{\partial \phi}
 + v_z \frac{\partial v_\varpi}{\partial z}
 - \frac{v^2_\phi}{\varpi}
  &=& - \frac{1}{\rho} \frac{\partial P}{\partial \varpi} 
 + f_\varpi \enspace , \label{eq:varpi_momentum}\\
   v_\varpi \frac{\partial v_\phi}{\partial \varpi}  
 + \frac{v_\phi}{\varpi} \frac{\partial v_\phi}{\partial \phi}
 + v_z \frac{\partial v_\phi}{\partial z}
 + \frac{v_\varpi v_\phi}{\varpi}
  &=& - \frac{1}{\rho \varpi} \frac{\partial P}{\partial \phi} 
 + f_\phi \enspace , \label{eq:phi_momentum}\\
   v_\varpi \frac{\partial v_z}{\partial \varpi}  
 + \frac{v_\phi}{\varpi} \frac{\partial v_z}{\partial \phi}
 + v_z \frac{\partial v_z}{\partial z}
  &=& - \frac{1}{\rho} \frac{\partial P}{\partial z} 
 + f_z \enspace ,\label{eq:z_momentum}
\end{eqnarray}
where $\rho$ is the gas mass density, $P$ is the pressure and $f_\varpi$, $f_\phi$ and $f_z$ are the components of the external forces acting on the gas.

Let us initially ignore any viscous effects (inviscid disc) and assume that the only force acting on the gas is gravity. If we further assume circular orbits --- $v_\varpi=0$, $v_\phi \neq 0$, and $v_z=0$, an assumption that will be droped later on when viscosity is included --- the only non-trivial fluid equations are the $\varpi$- and $z$-momentum equations (Eqs.~\ref{eq:varpi_momentum} and \ref{eq:z_momentum}), which take the form
\begin{eqnarray}
    \frac{1}{\rho} \frac{\partial P}{\partial \varpi} 
          &=& \frac{v^2_\phi}{\varpi} + f_\varpi \enspace , 
                                \label{eq:varpi_hydrostatic} \\
    \frac{1}{\rho} \frac{\partial P}{\partial z} 
          &=& f_z \enspace , 
                                \label{eq:z_hydrostatic}
\end{eqnarray}
where the external force components are given by the gravity of the spherical central star
\begin{eqnarray}
    f_\varpi &=& -\frac{GM\varpi}{(\varpi^2+z^2)^{3/2}} \enspace , \\
    f_z      &=& -\frac{GM z}{(\varpi^2+z^2)^{3/2}}     \enspace .
\end{eqnarray}

To specify the pressure, we introduce the equation of state, $P=c_s^2\rho$, where $c_s=(kT)^{1/2}(\mu m_{\rm H})^{-1/2}$. In this last expression, $k$ is the Boltzmann constant, $\mu$ is the gas molecular weight and $m_{\rm H}$ is the mass of the hydrogen atom.

In the thin-disc limit ($z \ll \varpi$), we obtain
\begin{eqnarray}
    v_\phi &=& V_{\rm crit} \left({R/\varpi}\right)^{1/2} \enspace , \label{eq:vphi}\\
       \frac{\partial \ln (c_s^2\rho)}{\partial z} 
  &=& -\frac{V^2_{\rm crit}Rz}{c_s^2 \varpi^3}
       \enspace . \label{eq:hseq}
\end{eqnarray}

The above equations mean that the disc rotates at the Keplerian orbital speed and is hydrostatically supported in the vertical direction. To determine the vertical disc structure we must solve Eq.~(\ref{eq:hseq}). Assuming an isothermal disc one obtains 
\begin{equation}
\rho(\varpi,z) = \rho_0(\varpi) \exp\left[ -0.5 (z/H)^2 \right] \, ,
\end{equation}
where $\rho_0$ the disc density at the mid-plane ($z = 0$), and the disc scale height 
is given by 
\begin{equation}
H(\varpi) = (c_s/v_\phi)\varpi \,.
\label{eq:scaleheight}
\end{equation}
Since $v_\phi \propto \varpi^{-0.5}$, we obtain the familiar result that  for an isothermal disc the scaleheight grows with distance from the star as $H \propto \varpi^{1.5}$.

As we shall see below, it is useful to express $\rho(\varpi,z)$ in terms of the disc surface density, 
$\Sigma$, written as
\begin{equation}
    \Sigma(\varpi)=\int_{-\infty}^\infty \rho(\varpi,z) \, dz = \sqrt{2\pi}H\rho_0 \enspace .
    \label{eq:sigmadef}
\end{equation}
Thus
\begin{equation}
\rho(\varpi,z) = \frac{\Sigma(\varpi)}{\sqrt{2\pi}H(\varpi)} \exp\left[ -0.5 (z/H)^2 \right] \, .
\label{eq:rho}
\end{equation}

\subsection{Viscous Outflow} 

Clearly, in Eq.~(\ref{eq:rho}) the disc density scale, $\Sigma(\varpi)$, is completely undetermined, because for a inviscid Keplerian disc we can choose to put an arbitrary amount of material at a given radius.
To set the density scale, we must include a mechanism --- viscous diffusion --- to transport material from the star outwards.

Viscous flows grow in a viscous diffusion timescale 
\begin{equation}
\tau_{\rm diff} = \varpi^2 / \nu \, ,
\end{equation} 
where $\nu$ is the kinematic viscosity. One problem that has already been noted long ago (\cite[Shakura \& Sunyaev 1973]{sha73}) is that for molecular 
viscosity the diffusion timescale is much too long. \cite{sha73} 
appealed instead to the so-called eddy (or turbulent) viscosity, which they parameterized as
\begin{equation}
\nu = \alpha c_s H \, ,
\end{equation}
where $0<\alpha<1$. The $\alpha$ parameter describes the ratio of the product of the 
turbulent eddy size and speed to the product of disc scaleheight 
and sound speed.  In other words, it is assumed that the largest eddies can be 
at most about the size of the disc scaleheight and that the ÒturnoverÓ velocity of the eddies cannot be arger than the sound speed (otherwise, 
the turbulence would be supersonic and the eddies would fragment into a series of 
shocks). 
With this value of the viscosity, the viscous diffusion timescale becomes 
\begin{eqnarray}
\tau_{\rm diff} &=& \frac{V_{\rm crit}}{\alpha c_s^2}\sqrt{\varpi R} \\
                         &\approx& 20 {\rm yr} \left(  \frac{0.01}{\alpha}  \right) \sqrt{\frac{\varpi}{R}} \, .
\end{eqnarray}
Studies of the formation and dissipation of the discs around Be stars find typical time scales of months to a few years (e.g., \cite[Wisniewski et al. 2010]{wis10}); therefore, an $\alpha$ of the order of 0.1 or larger is required to match the observed timescales (see Sect.~\ref{nonconstant}).

If we add viscosity to the fluid equations, we can still assume that the disc is axisymmetric and that the vertical structure is  hydrostatic ($v_z=0$).
However, the presence of an outflow implies that $v_\varpi \ne 0$. 
The $\varpi$- and $z$-momentum equations are the same as before, so $v_\phi$ 
and $\rho$ are the same as in the pure Keplerian case [eqs. (\ref{eq:vphi}) and (\ref{eq:rho})].

Two fluid equations remain to be solved, the continuity equation, Eq.~(\ref{eq:continuity}), and the $\phi$-momentum equation, Eq.~(\ref{eq:phi_momentum}). The continuity equation, written in terms of the surface density, 
\begin{equation}
 \frac{\partial}{\partial \varpi} (2 \pi \varpi \Sigma v_\varpi) = 0
                                                                \enspace                                                                
\label{eq:disk_continuity}
\end{equation}
means that the mass decretion rate, ${\dot M} \equiv 2 \pi \varpi \Sigma v_\varpi$, is a
constant (independent of $\varpi$).  The viscous outflow speed is 
given by
\begin{equation}
            v_\varpi = \frac{\dot M}{2 \pi \varpi \Sigma} \enspace .
\label{eq:radvel}
\end{equation}

This $\phi$-momentum equation now is more complicated because 
viscosity exerts a torque, which is described by the viscous shear stress 
tensor, $\pi_{\varpi\phi}$.  Including this shear stress, the $\phi$-momentum 
equation becomes
\begin{equation}
  v_\varpi \frac{\partial v_\phi}{\partial \varpi} 
     + \frac{v_\varpi v_\phi}{\varpi}
   = \frac{1}{\rho \varpi^2}
     \frac{\partial}{\partial \varpi} (\varpi^2 \pi_{\varpi\phi}) \enspace 
,
\label{eq:navier_stokes_phi_momentum}
\end{equation}
where
\begin{equation}
    \pi_{\varpi \phi} = \nu \rho \varpi 
          \frac{\partial (v_\phi/\varpi)}{\partial \varpi}
          = -\frac{3}{2}\alpha c_s^2 \rho
     \enspace .
\label{eq:shear_stress}
 \end{equation}
Multiplying Eq.~(\ref{eq:navier_stokes_phi_momentum}) by $\rho \varpi^2$ and 
integrating over $\phi$ and $z$, we find
\begin{equation}
    {\dot M} \frac{\partial}{\partial \varpi}(\varpi v_\phi)
    =   \frac{\partial}{\partial \varpi}(\mathcal{T})
  \enspace ,
\label{eq:jdot_gradient}
\end{equation}
where
\begin{equation}
\mathcal{T} = \int_{-\infty}^{\infty} \varpi \pi_{\varpi\phi} 2\pi \varpi dz = -3\pi \alpha c_s^2 \varpi^2\Sigma
\label{eq:torque1}
\end{equation}
is the viscous torque.
The $\phi$-momentum equation, Eq.~(\ref{eq:navier_stokes_phi_momentum}), expresses the fact that the change in the angular momentum flux --- $\varpi v_\phi$ being the specific angular momentum --- is given by the gradient of the viscous torque. Since the continuity equation  implies that ú$\dot{M}$ is constant, we integrate Eq.~(\ref{eq:navier_stokes_phi_momentum}) over $\varpi$ to obtain
\begin{equation}
\mathcal{T}(\varpi) = \dot{M} V_{\rm crit} \sqrt{\varpi R} + \mbox{constant} \, .
\label{eq:torque2}
\end{equation}
Substituting Eq.~(\ref{eq:torque1}) into Eq.~(\ref{eq:torque2}) and solving for the surface density we find
\begin{equation}
     \Sigma(\varpi)=\frac{\dot M}{3 \pi \alpha c_\mathrm{s}^2 }
            \left( \frac{G M}{\varpi^{3}} \right)^{1/2}
            \left[(R_0/\varpi)^{1/2}-1\right] \enspace .
\label{eq:disk_Sigma}
\end{equation}
Eq.~(\ref{eq:disk_Sigma}) describes the surface density of an unbounded disc (i.e. a disc that is allowed to grow indefinitely). The integration constant $R_0$ is a parameter that depends on the integration constant of Eq.~(\ref{eq:torque2}) and is related to the physical size of the disc;
for time-dependent models, such as those of \cite{oka07}, $R_0$ grows with time and thus $R_0$ is related with the age  of the disc.

\subsection{Properties of the Solution}

We have now completed the hydrodynamic description of an isothermal, unbounded viscous decretion disc. To fully determine the problem one must specify the decretion rate, $\dot{M}$, the value of $\alpha$, and the disc age (or size). Assuming that the disc is sufficiently old, $R_0 \gg R$, in which case Eq.~(\ref{eq:disk_Sigma}) becomes a simple power-law with radius, $\Sigma(\varpi)\propto \varpi^{-2}$. 
From Eqs.~(\ref{eq:rho}) and (\ref{eq:scaleheight}) we obtain that the isothermal disc density profile is quite steep, $\rho \propto \varpi^{-3.5}$.

Another important property of isothermal viscous discs can be readily derived from Eq. (\ref{eq:scaleheight}). Since $v_\phi$ is much larger than the sound speed (the former is of the order of several hundreds of km/s whereas the later is a few tens of km/s), the disc scaleheight is small compared to the stellar radius, i.e., the disc is geometrically thin.
Finally, from Eq.~(\ref{eq:radvel}) we find that, for large discs,  the radial velocity is a
linear function of the radial distance, $v_\varpi\propto\varpi$.

   \begin{figure}
   \centering
   \includegraphics*[width=2.6in,height=1.6in]{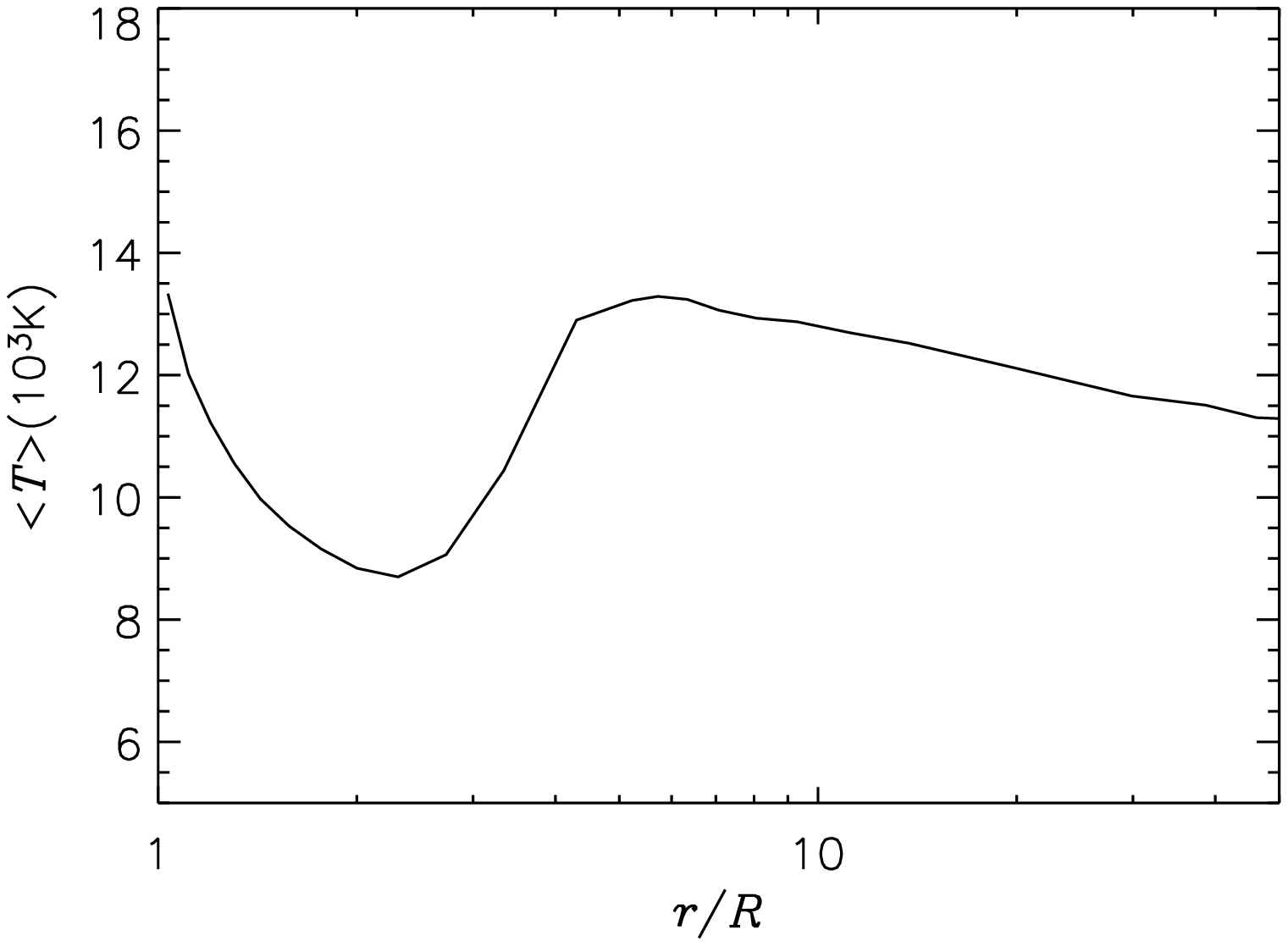}
   \includegraphics*[width=2.6in,height=1.6in]{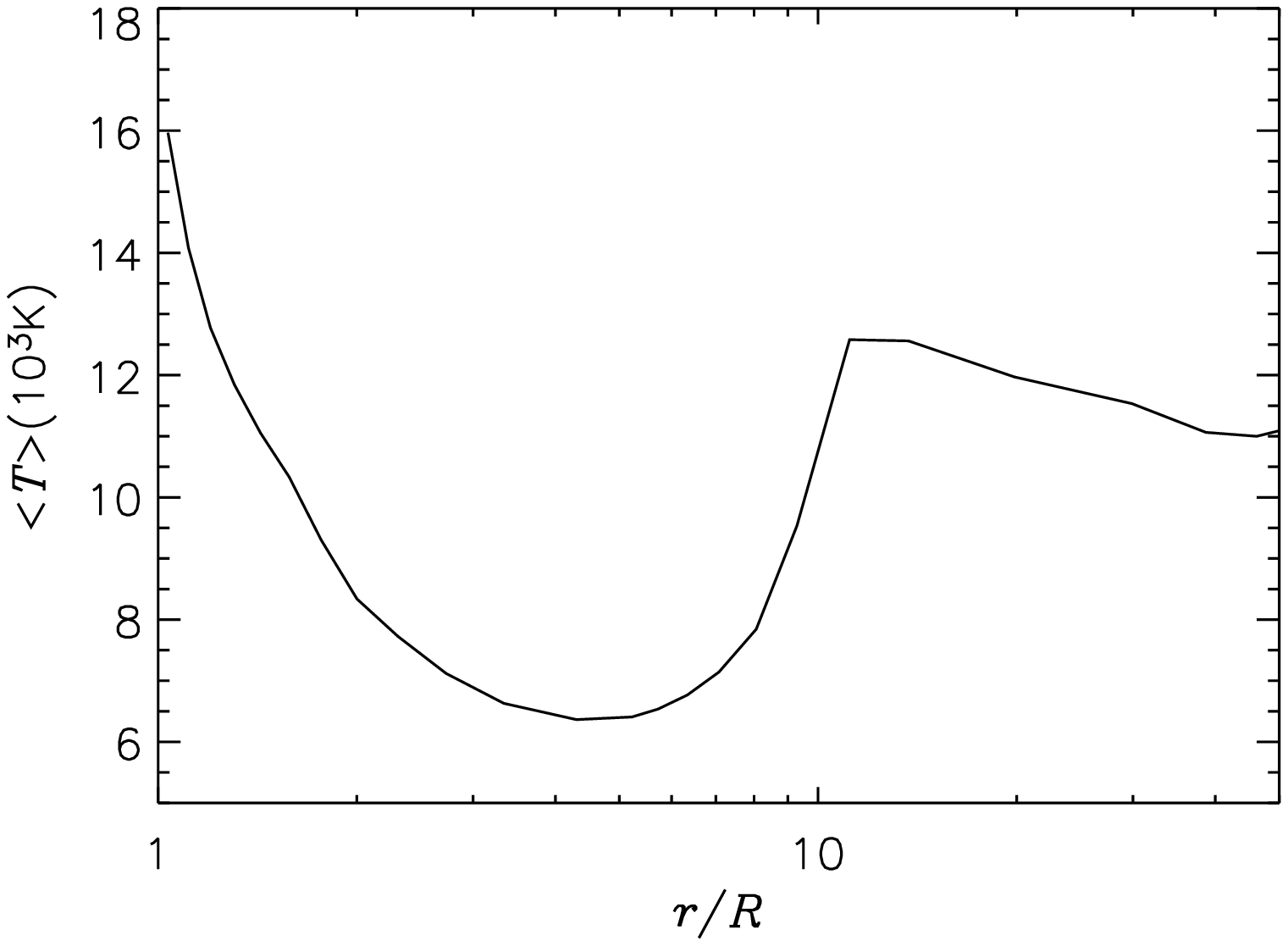}
   \includegraphics*[width=2.6in,height=1.6in]{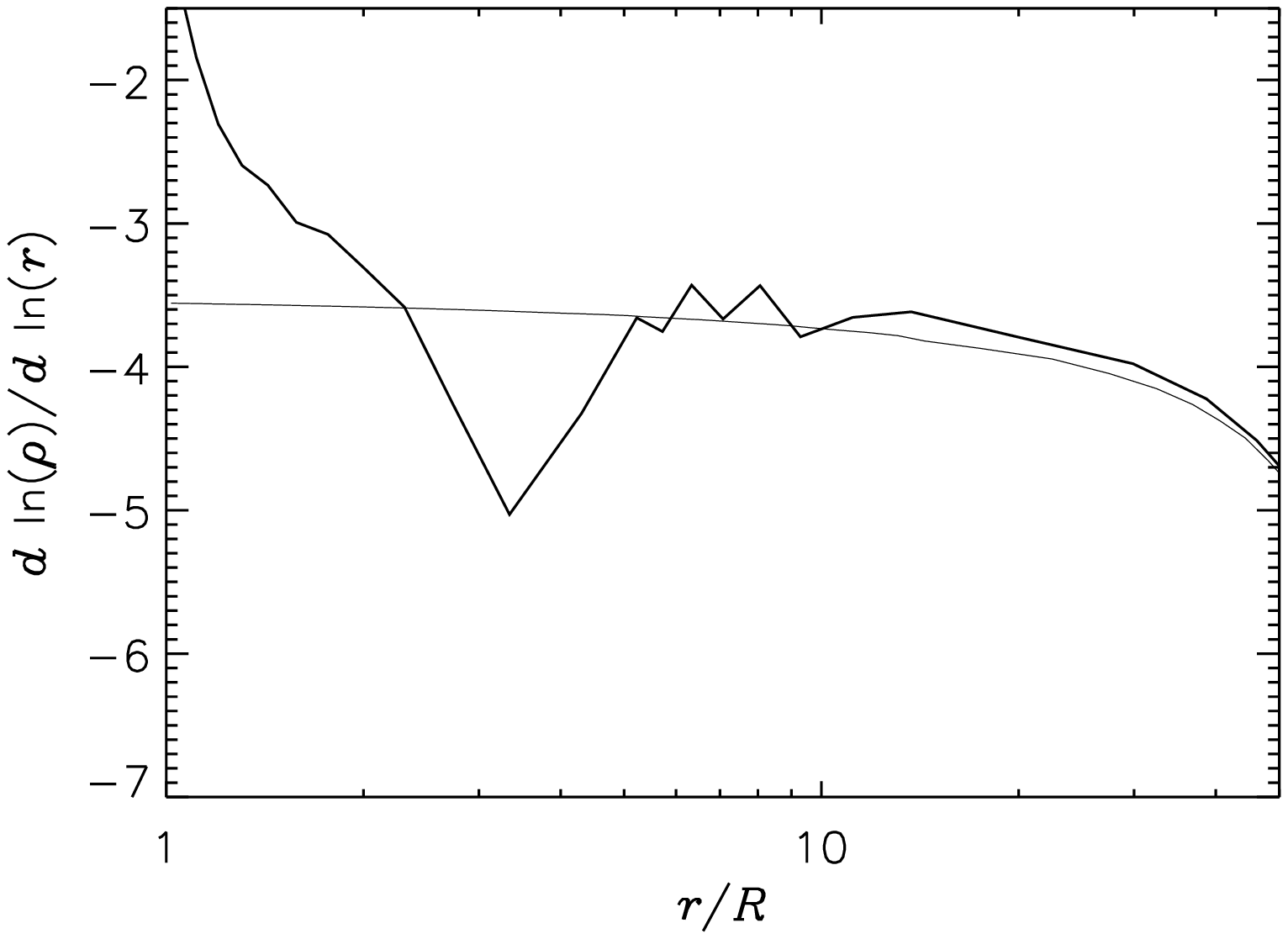}
    \includegraphics*[width=2.6in,height=1.6in]{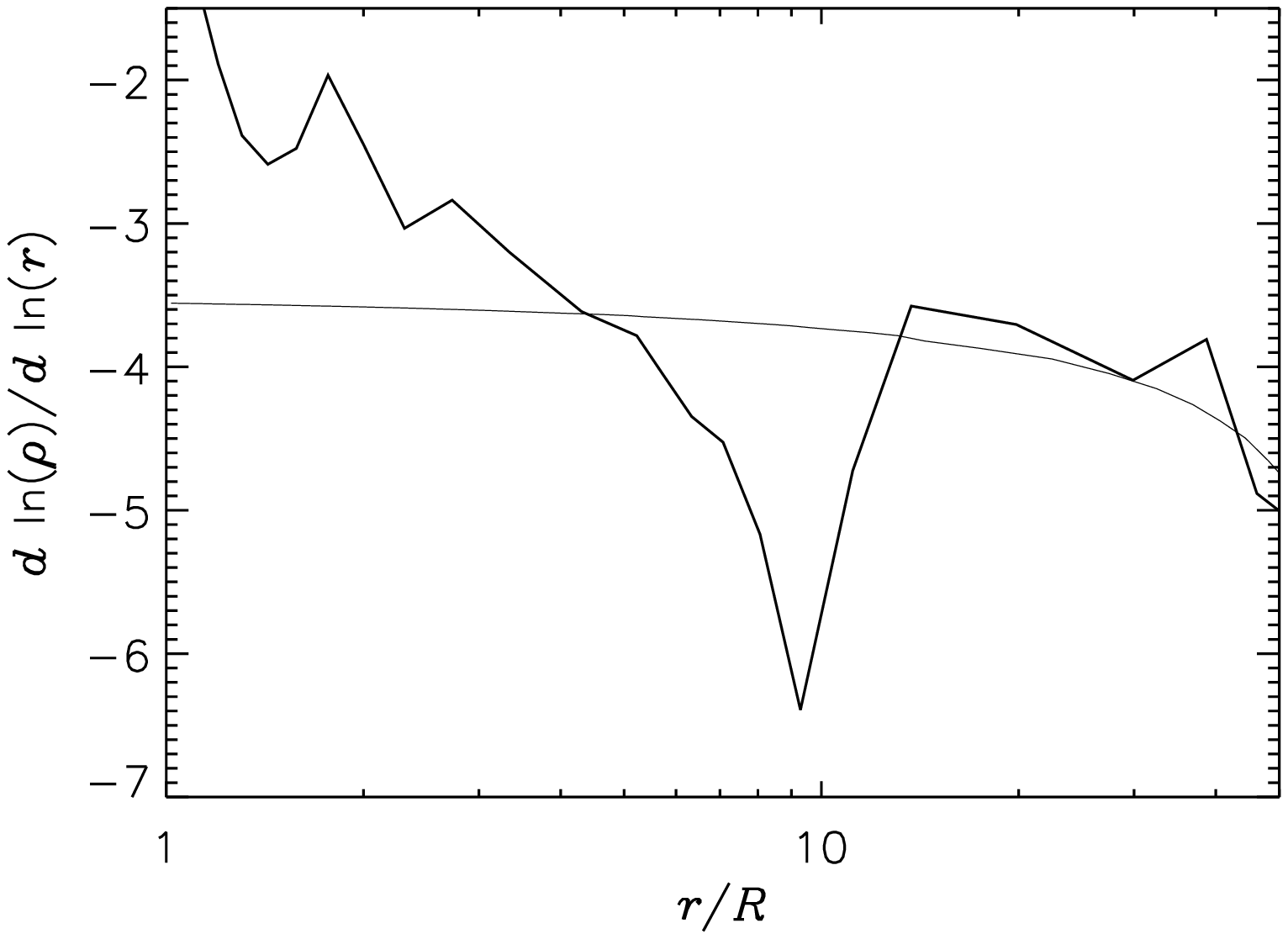}
      \caption{
      \emph{Top}: Vertically averaged disc temperature.
      \emph{Bottom}: Local power-law index of the density profile.
      \emph{Left: } Results for a low-density model ($\rho_0 = 4.2\times 10^{-12} \rm\; g\;cm^{-3}$).
      \emph{Right: } Results for a high-density model ($\rho_0 = 3.0\times 10^{-11} \rm\; g\;cm^{-3}$). 
   Adapted from \cite{car08}.
        }
    \label{fig:temperature}
   \end{figure}

\subsection{Temperature structure}

The temperature structure of viscous discs were investigated by \cite{car06a} and \cite{sig07}. They found that those discs are highly nonisothermal, mainly on their denser inner parts. 
An example of the temperature structure is shown in Fig.~\ref{fig:temperature}. The temperature initially drops very quickly close to the stellar photosphere; when the disc becomes optically thin 
vertically the temperature rises back to the 
optically thin radiative equilibrium temperature, which is approximately
constant, as in the winds of hot stars.  The density is the most important factor that controls the temperature structure: for high density models the amplitude of the temperature variations is much larger and the nonisothermal region extends much farther out into the disc.

\subsection{Non-isothermal effects on the disc structure}

From Eqs.~(\ref{eq:rho}) and (\ref{eq:disk_Sigma}) we see that both the viscous diffusion and the vertical hydrostatic solutions depend on the gas temperature; one can expect, therefore, that the complex temperature structure of the disc might have effects on the disc density structure. \cite{car08} and \cite{sig09} calculated the vertical density structure of  a disc in consistent vertical hydrostatic equilibrium. They found that the temperature decrease causes the disc to collapse, becoming much thinner in the inner regions.  This collapse redistributes the disc material toward the equator, increasing the midplane density by a factor of up to 3 relative to an equivalent isothermal model.

\cite{car08} investigated, in addition, how the temperature affects the viscous diffusion.
The combination of the radial temperature structure, disc scaleheight, and viscous transport produces a complex radial dependence for the disc density that departs very much from the simple $n=3.5$ power-law. As shown in Fig.~\ref{fig:temperature}, the equivalent radial density exponent varies between $n=2$ in the inner disc to $n=5$ near the temperature minimum, eventually rising back to the isothermal value $n=3.5$ in the outer disc.
We conclude that non-isothermal effects on the viscous diffusion may account for the at least part of the large scatter of the index $n$ reported in the literature (Sect.~\ref{density}).

\subsection{Two test cases: $\zeta$ Tau and $\chi$ Oph}

The model described above makes several predictions about the disc structure that are in qualitative agreement with the observations (Sect.~\ref{observations}): viscous discs are geometrically thin, rotate in a near keplerian fashion ($v_\phi \propto \varpi^{-1/2}$ and $v_\varpi \ll v_\phi$) and have a steep density fall-off ($n=-3.5$ in the isothermal case).
This model and its predictions must now be quantitatively compared to observations.

\begin{figure}
   \centering
   \includegraphics*[width=2.5in,viewport=0 200 170 299]{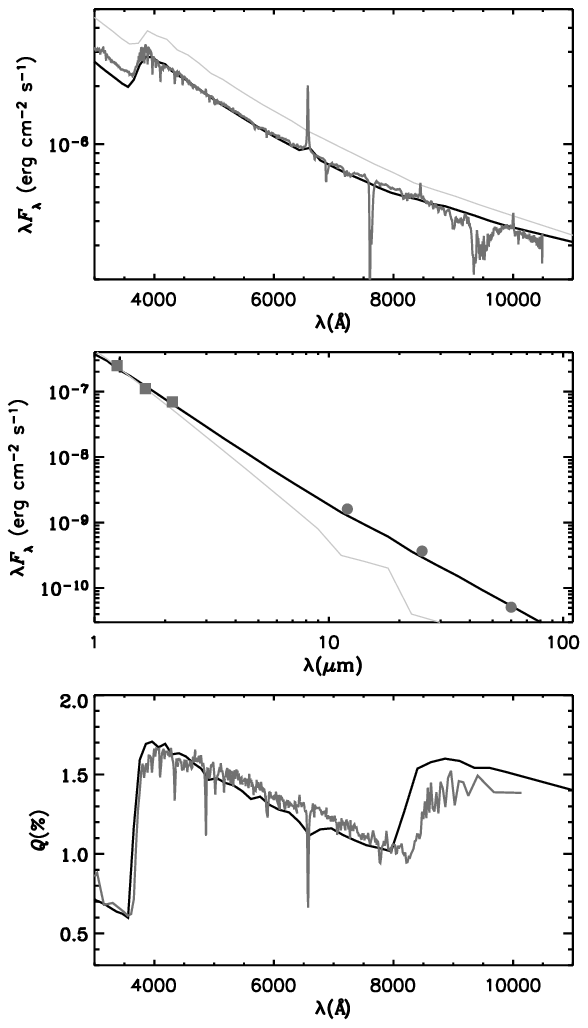}
   \includegraphics*[width=2.5in,viewport=0 0 170 100]{review_carciofi_fig4.eps}
      \caption{\emph{Left: } Fit to the SED of $\zeta$ Tau. The dark grey lines are the observations and {the black lines the model results}. The light grey line corresponds to the unattenuated stellar SED and gives a measure of how the disc affects the emergent flux. Adapted from \cite{car09}.
        }
    \label{fig:zetatau}
   \end{figure}

A successful verification of the viscous decretion disc model have been obtained in the case of the Be star $\zeta$ Tau (\cite[Carciofi et al. 2009]{car09}). This star is particularly suitable for such study because it had shown little or no secular evolution in the past 18 years or so (\cite[{\v S}tefl et al. 2009]{stefl09}); therefore, a constant mass decretion rate, as assumed above when deriving the disc structure, is a good approximation for this system.
Some of the results obtained for $\zeta$ Tau are shown in Fig.~\ref{fig:zetatau}.
The radial dependence of the disc density, temperature, and opening angle all affect the slope of the visible and IR SED (\cite[Waters 1986]{wat86}, \cite[Carciofi \& Bjorkman 2006]{car06a}), as well as the shape of the intrinsic polarization.
Therefore, the fact that the model reproduces the detailed shapes of SED and spectropolarimetry represents a non-trivial test of the Keplerian decretion disc model. 

Using a somewhat different model, \cite{tyc08} successfully fitted several observations of the Be star $\chi$ Oph, including high angular resolution interferometry. In their modeling the index $n$ of the density power-law is a free parameter. Interestingly, their best fitting model was for a much flatter density law ($n=2.5$). Whether this flatter density profile can be accommodated by including some other physical process in addition to viscosity (e.g., tidal effects by a binary) or is an indication that the viscous decretion disc model is not a good model for this system remains to be verified by further analysis.

\section{Viscous Discs Fed by Non-constant Decretion Rates \label{nonconstant}}

All the discussion so far have been focused on the problem of a disc fed by a constant decretion rate that grows to a given size in a viscous timescale. 
Clearly, those models can only be applied to objects, such as $\zeta$ Tau, that went through a sufficiently long and stable decretion phase.
A different approach is needed if one wants to investigate dynamically active systems such as the one shown in Fig.~\ref{dewit}.

\begin{figure}
   \centering
   \includegraphics*[width=5.2in]{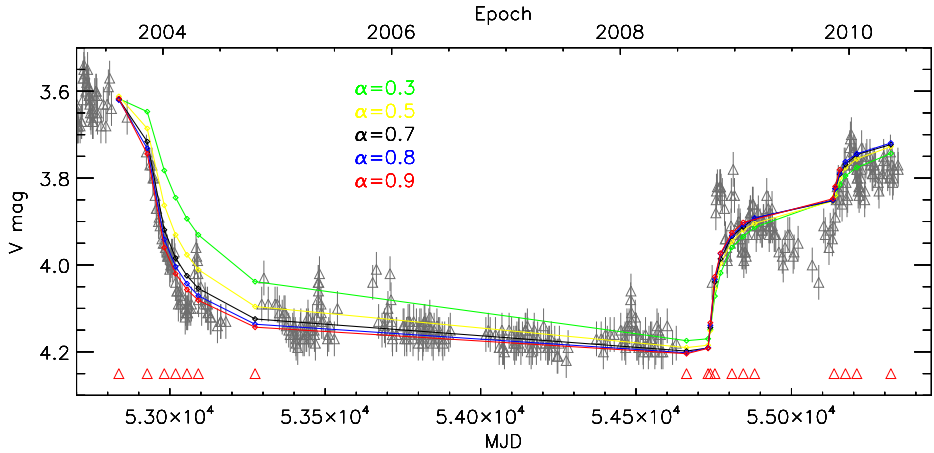}
      \caption{The visual light curve of 28 CMa fitted with a dynamical viscous decretion disc model with different values of the viscosity parameter, $\alpha$ (\cite[\v{S}tefl et al. 2010]{stefl10}) }
    \label{28cma}
   \end{figure}

\cite{oka07}, \cite{hau10} and \cite{jon08} described the solution of the viscous diffusion problem for systems with non-constant mass decretion rates. 
Fig.~\ref{28cma} shows a series of models of the lightcurve of the Be star 28 CMa (Carciofi et al., in prep.), that uses the computer code {\sc singlebe} by Atsuo Okazaki to solve the 1D viscous diffusion problem and the {\sc hdust} code to calculate the emergent spectrum (\cite[Haubois et al. 2010]{hau10}).  The simulation begins after a few years of disc evolution to account for the previous disc build-up. At $t=2003.6$ mass decretion is turned off and the system evolves passively until $t=2008.8$, when the recent outburst started (\cite[\v{S}tefl et al. 2010]{stefl10}). The model with $\alpha=0.9$ is the one that reproduces best the lightcurve at all phases. This is, to our knowledge, the first time the viscosity parameter is determined for a Be star disc.

\section{conclusions}

This paper discussed the basic hydrodynamics that determine the structure of viscous decretion disc models. 
Those are, to date, the most satisfactory models for Be discs because they can account quantitatively for most of their observational properties, namely the fact that they are geometrically thin, have Keplerian rotation and steep radial density profiles.
It has been recently shown that the viscous decretion disc model can also explain the temporal evolution of dynamically active systems such as 28 CMa. Studies of such system allows for the determination of the viscosity parameter $\alpha$ and the mass decretion rate, $\dot{M}$, two quantities that are very difficult to determine otherwise.

\acknowledgements{The author acknowledges support from FAPESP grant 2010/08809-5 and CNPq grant 308985/2009-5.}

\begin{discussion}

\discuss{Puls}{Excellent talk! How are the occupation numbers required for the profile synthesis calculated?}

\discuss{Carciofi}{The computer code {\sc hdust} calculates the occupation numbers by solving the coupled problem of the radiative transfer, radiative equilibrium and statistical equilibrium in the NLTE regime. Because the code uses the Monte Carlo method, it can handle arbitrary 3D density and velocity distributions.}

\discuss{Lee}{What is the timescale of the $V/R$ variability?}

\discuss{Carciofi}{If the question refers to cyclic $V/R$ variability, such as observed in $\zeta$ Tau, the timescale is of a few years}

\end{discussion}


\begin{thebibliography}{}

\bibitem[Bjorkman 
\& Cassinelli (1993)]{bjo93} Bjorkman, J.~E., \& Cassinelli, J.~P.\ 1993,  \textit{ApJ}, 409, 429 

\bibitem[Bjorkman (1997)]{bjo97} Bjorkman, J.~E.\ 1997, Circumstellar Disks, in Stellar Atmospheres: Theory and Observations, ed. J.~P.~de~Greve, R.~Blomme, 
\& H.~Hensberge, (New York: Springer)

\bibitem[Bjorkman (2000)]{bjo00} Bjorkman, J.~E.\ 2000, IAU 
Colloq.~175: The Be Phenomenon in Early-Type Stars, 214, 435 

\bibitem[Bjorkman \& Carciofi (2005)]{bjo05} Bjorkman, J.~E., \& Carciofi, A.~C. 2005,  in ASP Conf. Ser. 337, The Nature and Evolution of Disks Around Hot Stars, ed. R. Ignace \& K. G. Gayley (San 
Francisco: ASP), 75

\bibitem[Carciofi \& Bjorkman (2006)]{car06a} Carciofi, A.~C., 
\& Bjorkman, J.~E.\ 2006, \textit{ApJ}, 639, 1081 

\bibitem[Carciofi 
\& Bjorkman (2008)]{car08} Carciofi, A.~C., \& Bjorkman, J.~E.\ 2008, \textit{ApJ}, 684, 1374 

\bibitem[Carciofi et 
al. (2009)]{car09} Carciofi, A.~C., Okazaki, A.~T., Le Bouquin, J.-B., {\v S}tefl, S., Rivinius, T., Baade, D., Bjorkman, J.~E., \& Hummel, C.~A.\ 2009, \textit{A\&A}, 504, 915 

\bibitem[Clark et 
al. (2003)]{cla03} Clark, J.~S., Tarasov, A.~E., \& Panko, E.~A.\ 2003, \textit{A\&A}, 403, 239 

\bibitem[Draper et al. (2010)]{dra10} Draper, Z. H. et al. 2010, these proceedings

\bibitem[Jones et al. (2008)]{jon08} Jones, C.~E., Sigut, 
T.~A.~A., \& Porter, J.~M.\ 2008, \textit{MNRAS}, 386, 1922 

\bibitem[Hanuschik (1996)]{han96} Hanuschik, R.~W.\ 1996, \textit{A\&A}, 308, 170 

\bibitem[Harmanec(1983)]{har83} Harmanec, P.\ 1983, Hvar 
Observatory Bulletin, 7, 55 

\bibitem[Haubois et al. (2010)]{hau10} Haubois, X., Carciofi, A. C. \& Okazaki, A. T. Ph.\ 2010, these proceedings

\bibitem[Kato (1983)]{kat83} Kato, S.\ 1983, \textit{PASJ}, 35, 249

\bibitem[Lee et al. (1991)]{lee91}
Lee, U., Saio, H., Osaki, Y.\ 1991,  \textit{MNRAS}, 250, 432

\bibitem[Meilland et al. (2007a)]{mei07a} Meilland, A., et al.\ 2007, \textit{A\&A}, 464, 59 

\bibitem[Meilland et al. (2007b)]{mei07b} Meilland, A., et al.\ 2007, \textit{A\&A}, 464, 73 

\bibitem[Okazaki (1991)]{oka91} Okazaki, A.~T.\ 1991, \textit{PASJ}, 43, 75

\bibitem[Okazaki (1997)]{oka97} Okazaki, A.~T.\ 1997, \textit{A\&A}, 318, 548 

\bibitem[Okazaki (2001)]{oka01} Okazaki, A.~T.\ 2001, \textit{PASJ}, 53, 119 

\bibitem[Okazaki et al. (2002)]{oka02} Okazaki, A.~T., Bate, 
M.~R., Ogilvie, G.~I., \& Pringle, J.~E.\ 2002, \textit{MNRAS}, 337, 967

\bibitem[Okazaki (2007)]{oka07} Okazaki, A.~T.\ 2007, Active 
OB-Stars: Laboratories for Stellar and Circumstellar Physics, 361, 230 

\bibitem[Oudmaijer et al. (2010)]{oud10} Oudmaijer, R. et al. 2010, these proceedings

\bibitem[Porter (1999)]{por99} Porter, J.~M. 1999,  \textit{A\&A}, 348, 512

\bibitem[Porter \& Rivinius (2003)]{por03} Porter, J.~M., \& Rivinius, T.\ 2003, \textit{PASP}, 115, 1153 

\bibitem[Pringle (1981)]{pri81} Pringle, J.~E.\ 1981, \textit{ARAA}, 19, 137 

\bibitem[Quirrenbach et al. (1997)]{qui97} Quirrenbach, A., et. al.  1997, \textit{ApJ}, 479, 477

\bibitem[Rivinius et 
al. (1998)]{riv98} Rivinius, T., Baade, D., Stefl, S., Stahl, O., Wolf, B., \& Kaufer, A.\ 1998, \textit{A\&A}, 333, 125 

\bibitem[Rivinius (2007)]{riv07} Rivinius, T.\ 2007, Active 
OB-Stars: Laboratories for Stellare and Circumstellar Physics, 361, 219

\bibitem[Shakura \& Sunyaev (1973)]{sha73}
  Shakura N.~I., Sunyaev R.~A., 1973, \textit{A\&A}, 24, 337

\bibitem[Stee (2010)]{ste10} Stee, Ph.\ 2010, these proceedings

\bibitem[Sigut \& Jones (2007)]{sig07} Sigut, T.~A.~A., \& Jones, C.~E.\ 2007, \textit{ApJ}, 668, 481 

\bibitem[Sigut et al. (2009)]{sig09} Sigut, T.~A.~A., McGill, 
M.~A., \& Jones, C.~E.\ 2009,  \textit{ApJ}, 699, 1973

\bibitem[{\v S}tefl et al.(2009)]{stefl09} {\v S}tefl, S., et al.\ 2009,  \textit{A\&A}, 504, 929 

\bibitem[\v{S}tefl et al. (2010)]{stefl10} \v{S}tefl , S. et al. 2010, these proceedings

\bibitem[Townsend et al. (2005)]{tow05} Townsend, R.~H.~D., 
Owocki, S.~P., \& Groote, D.\ 2005, \textit{ApJ}, 630, L81 

\bibitem[Tycner et al. (2008)]{tyc08} Tycner, C., Jones, 
C.~E., Sigut, T.~A.~A., Schmitt, H.~R., Benson, J.~A., Hutter, D.~J., 
\& Zavala, R.~T.\ 2008, \textit{ApJ}, 689, 461 

\bibitem[Waters (1986)]{wat86} Waters, L.~B.~F.~M.\ 1986,  \textit{A\&A}, 162, 121 

\bibitem[Wisniewski et al. (2010)]{wis10} Wisniewski, J.~P., 
Draper, Z.~H., Bjorkman, K.~S., Meade, M.~R., Bjorkman, J.~E., 
\& Kowalski, A.~F.\ 2010, \textit{ApJ}, 709, 1306

\bibitem[de Wit et 
al. (2006)]{dew06} de Wit, W.~J., Lamers, H.~J.~G.~L.~M., Marquette, J.~B., \& Beaulieu, J.~P.\ 2006, \textit{A\&A}, 456, 1027 

\bibitem[Wood, Bjorkman, \& Bjorkman (1997)]{woo97} Wood, K., Bjorkman, K.~S., \& Bjorkman, J.~E.\ 1997, \textit{ApJ}, 477, 926

\end{thebibliography}
\end{document}